\documentclass[twocolumn,showpacs,preprintnumbers,nofootinbib,
               superscriptaddress]{revtex4}
\usepackage{graphicx}

\begin{document}

\preprint{BNL-NT-07/1}

%%%%%%%%%%%%%%%%%%%%%%%%%%%%%%%%%%%%%%%%%%%%%%%%%%%%%%%%%%%%%%%%%%%%%%%

\title{Constant contribution in meson correlators 
at finite temperature} 

%%%%%%%%%%%%%%%%%%%%%%%%%%%%%%%%%%%%%%%%%%%%%%%%%%%%%%%%%%%%%%%%%%%%%%%

\author{Takashi Umeda\\
Physics Department, Brookhaven National Laboratory,Upton, NY 11973, USA} 

%\footnote{Present address: University of Tsukuba,
%Tsukuba, Ibaraki 305-8571, Japan.}}
%\affiliation{Physics Department, Brookhaven National Laboratory,
%Upton, NY 11973, USA} 

\date{\today}


\begin{abstract}
We discuss a constant contribution to meson correlators at finite
temperature. In the deconfinement phase of QCD,
a colored single quark state is allowed as a finite energy state,
which yields to a contribution of wraparound quark propagation to
temporal meson correlators. We investigate the effects in the free quark
case and quenched QCD at finite temperature. The ``scattering''
contribution causes a constant mode in meson correlators with zero
spatial momentum and degenerate quark masses, which can dominate the
correlators in the region of large imaginary times. 
In the free spectral function, the contribution yields a term
proportional to $\omega\delta(\omega)$. Therefore this contribution is
related to transport phenomena in the quark gluon plasma. 
It is possible to distinguish the constant contribution from the other 
part using several analysis methods proposed in this paper. 
As a result of the analyses, we find that drastic changes in
charmonium correlators for $\chi_c$ states just above the deconfinement
transition are due to the constant contribution. The other differences
in the $\chi_c$ states are small. It may indicate the survival of
$\chi_c$ states after the deconfinement transition until, at least,
$1.4T_c$. 
\end{abstract}

\maketitle

\section{Introduction}
\label{sec:Introduction}
In the imaginary time formalism of finite temperature field theory,
temporal correlators are related to several dynamical properties
through a prescription of analytical continuation. 
Especially the imaginary time correlator can be described by 
the same spectral function as that of retarded and advanced green
functions. 
Therefore the spectral functions extracted from the imaginary time
correlators, which are able to be calculated in lattice QCD, enable
us to investigate important physics at finite temperature, 
e.g. dilepton rates \cite{Karsch:2001uw}, light mesons
\cite{deForcrand:2000jx}, heavy quarkonia
\cite{Umeda:2000ym,Umeda:2002vr,Datta:2003ww,Asakawa:2003re,
Aarts:2006nr,Jakovac:2006sf}, 
glueballs \cite{Ishii:2002ww}, transport coefficients 
\cite{Aarts:2002cc,Gupta:2003zh,Nakamura:2004sy}.

One of the most important quantities at finite temperature QCD is the
spectral function of heavy quarkonium, which plays the key role for
understanding the quark gluon plasma (QGP) formation in heavy ion
collision experiments e.g. the RHIC experiment at Brookhaven National
Laboratory. 
Recent studies on the spectral function of charmonium above $T_c$
suggest that hadronic excitations corresponding to $J/\psi$ may survive
in the deconfinement phase till relatively high temperature
\cite{Umeda:2002vr,Datta:2003ww,Asakawa:2003re,Aarts:2006nr,Jakovac:2006sf}.
Such results of strongly interacting QGP may affect the  
scenario of $J/\psi$ suppression \cite{Hashimoto:1986nn,Matsui:1986dk}.
Therefore the determination of the accurate dissociation temperature
of $J/\psi$ is required for many phenomenological studies.

As somewhat different from the ``direct'' $J/\psi$ suppression,
recently an ``indirect'' $J/\psi$ suppression was proposed 
\cite{Digal:2001bh,Karsch:2005nk}.
The total yield of $J/\psi$ is not only from the direct production, 
about 40\% of the $J/\psi$'s that come from the higher states $\psi'$
and $\chi_c$ \cite{Antoniazzi:1992iv}. 
If the higher states dissociate at lower temperatures, 
a part of the $J/\psi$ suppression may be observed in experiments 
at a lower temperature than that of $J/\psi$. 
Therefore the study of $\psi'$ and $\chi_c$ states and 
their dissociation temperatures is a very interesting subject for the
physics of QGP.
Several groups have already investigated the $\chi_c$ states, and 
their results indicate that the lowest peak of the spectral function, 
corresponding to the $\chi_c$ state, may dissociate just after the
transition temperature \cite{Datta:2003ww,Aarts:2006nr,Jakovac:2006sf}.
On the other hand there is no lattice QCD
study on the $\psi'$ states above $T_c$,
because it is difficult to extract information of non-lowest states
such as $\psi'$ from the correlator on the lattice at finite temperature.

\begin{figure}
\resizebox{50mm}{45mm}{
 \includegraphics{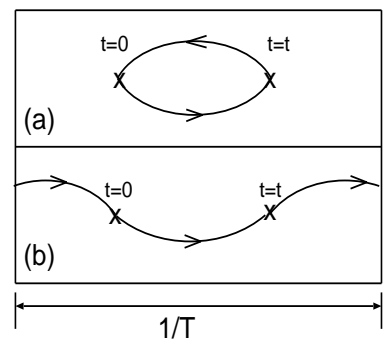}}
\caption{A sketch of quark line diagrams for a meson-like correlator 
in a system with finite temporal extent. 
Vertical lines show the boundaries in temporal direction.}
\label{fig:corr}
\end{figure}

In such studies of temporal correlators at finite temperature, 
one has to take account of a characteristic contribution caused
by the finite temporal extent.
Usually a meson correlator is interpreted by a diagram with quark and
anti-quark propagators like it is sketched in Fig.~\ref{fig:corr}(a).
However in the case of a system with a finite temporal extent, a
wraparound (scattering) contribution, as shown in
Fig.~\ref{fig:corr}(b), has to be included as well. 
It is similar to the situation at zero temperature 
e.g. the study of two pion correlators \cite{Kim:2003xt}, and pentaquark 
correlator \cite{Takahashi:2005uk}.
Because these correlators include multi-hadron states as intermediate
states, and each state is also allowed as a color singlet state, which
can have a finite energy state in the confined phase.
At zero temperature the calculation should be performed, in principle,
with infinite temporal extent, therefore, the wraparound effect is not
physical. The contribution is usually removed using e.g. 
Dirichlet boundary conditions \cite{Takahashi:2005uk} 
or appropriate analyses \cite{Kim:2003xt}.

In finite temperature calculations using the imaginary time formalism, 
the temporal extent is determined by (the inverse of) the temperature,
and the quark fields have to have anti-periodic boundary conditions.
In the confinement phase, an expectation value of the
diagram in Fig.\ref{fig:corr}(b) vanishes because it is related to a
single quark propagation, which is exponentially
suppressed with its infinite energy due to the confinement.
On the other hand, in the deconfinement phase, the contribution may
yield a finite expectation value.
This is similar to the discussion of the Polyakov loop expectation value.
For example in the free quark system, when one adopt a meson-like
operator with degenerate quark masses and zero spatial momentum,
the correlator of the diagram in Fig.\ref{fig:corr}(a) behaves like
$\exp(-m_qt)\times\exp(-m_qt)=\exp(-2m_qt)$, where $m_q$ is the quark
mass (or the energy of a single quark state). 
(Here we note that equations are written with dimension-less
variables throughout this paper, if there is no comment on the units.)  
On the other hand, the diagram in Fig.\ref{fig:corr}(b) behaves like
$\exp(-m_qt)\times\exp(-m_q(L_t-t))=\exp(-m_qL_t)$, where $L_t$ is the 
inverse of the temperature.  
The latter diagram provides a constant contribution to the correlator,
and the contribution may dominate in the large imaginary time region as
a zero energy mode. 
Due to the contribution, correlators of such meson-like operators,
e.g. charmonia, may be drastically changed just after the deconfinement
transition.
The latter type diagram, in general, yields t-dependent contributions
which depends on a difference between single quark energies for each
quark propagator. 
Even in non-degenerate quark masses or non-zero spatial momentum
cases, the diagram can yield the constant contribution when the
energies are identical. However we do not discuss the case in this
paper.  The finite momentum case has been discussed in
Ref.~\cite{Aarts:2005hg} at high temperature limit. 

This paper is organized as follows.
In Sect.~\ref{sec:FreeTheory} we discuss the correlators of meson-like 
operators and its spectral functions in a system of free quarks with
finite temporal extent.
Some numerical results are also presented.
In this section we also discuss the physical interpretation of the
constant contribution and some analysis methods to avoid this
contribution. 
In Sect.~\ref{sec:quench_qcd} we show numerical simulations in quenched
QCD on anisotropic lattices, and demonstrate effects of the constant 
contribution
in the interacting case and the analyses to avoid the contribution. 
As a result of these analyses we show that the $\chi_c$ state yields 
a small change except for the constant mode in its spectral function
after the deconfinement transition like the $J/\psi$ state
at least up to $T=1.4T_c$. 
At last we discuss results on the constant contribution.

\section{The constant contribution in the free theory}
\label{sec:FreeTheory}

\subsection{Meson-like correlators in the free quark case}
\label{sec:freecorr}

In order to investigate the constant contribution as mentioned in
the Introduction, 
we first consider the free quark case of QCD, in which the constant
contribution can be easily calculated.
Here we define the meson-like correlators with quark
bilinear operators,  
$O_{\Gamma}(\vec{x},t)=\bar{q}(\vec{x},t)\Gamma q(\vec{x},t)$,
in the free quark case this gives for the correlator,
\begin{eqnarray}
C(t)=\sum_{\vec{x}}\langle O_\Gamma(\vec{x},t)
O_\Gamma^\dagger(\vec{0},0) \rangle,
\label{eq:corrdef}
\end{eqnarray}
where $\Gamma$ are appropriate $4\times 4$ matrices, i.e.
$\gamma_5$, $\gamma_i$, $1$, and $\gamma_i\gamma_5$ for 
pseudoscalar(Ps), vector(V), scalar(S), and axialvector(Av) channels
respectively.
In this section and later we always adopt the anti-periodic boundary
condition for quark fields in temporal direction to suppose a
finite temperature case,  
and periodic boundary conditions in spatial directions. 
In this paper we calculate correlators for degenerate quark masses,
constructed from bilinear operators, with vanishing spatial momentum as
the simplest case, 
which is aiming at studies of charmonium at finite temperature
discussed in Sect.~\ref{sec:quench_qcd}.
The spectral function of the correlator is defined by
\begin{eqnarray}
C(t)&=&\int_0^\infty d\omega \rho_\Gamma(\omega)K(\omega,t), \nonumber\\
&&K(\omega,t)=\frac{\cosh\left(\omega(\frac{L_t}{2}-t)\right)}
{\sinh{\left(\omega\frac{L_t}{2}\right)}},
\label{eq:spf}
\end{eqnarray}
and has been calculated in Ref.~\cite{Karsch:2003wy,Aarts:2005hg} 
in the high temperature limit.

In the spectral representation, the constant contribution
provides a term proportional to $\omega\delta(\omega)$, which
corresponds to a constant mode in the meson-like correlators.
This term is an odd function in frequency $\omega$, which is one of the
basic properties of bosonic spectral function. 
From the results in \cite{Karsch:2003wy,Aarts:2005hg} 
one can find several characteristic properties of the constant
contribution.
First there is no constant mode in the Ps channel,
however the other correlators have the constant mode, which comes from
the scattering contribution.
The correlator in the S channel has also no constant mode at vanishing
quark mass, where the chiral symmetry is restored.
Secondly, in the V channel (and the Av channel in the chiral limit),
there is no constant contribution coming from the zero relative momentum
process. 
On the other hand, there is one in the Av and S channels (and it
disappears in chiral limit)\footnote{One can find this property from
the Table I of Ref.~\cite{Karsch:2003wy}. No constant mode and zero
relative momentum correspond to $d^{lat}_H=0$ and $d=0$ (if quark mass
is finite) respectively.}.   
These properties are very important especially in a finite volume
calculation. When the minimum nonzero momentum due to the finite volume
is large, the contribution of the zero relative momentum processes
dominates in the correlators. 
Furthermore, a two quark state with zero relative
momentum is forbidden in the P-wave state.
Therefore the relative contribution of the constant mode increases
exponentially as the spatial volume decreases in the P-wave state at
finite quark mass. 
As a result, we can expect a very different behavior of correlators
between S-wave and P-wave states in the case of finite quark mass and
finite (but relatively small) volume. 
Thirdly, correlators for each channel coincide at the midpoint $t=L_t/2$
as mentioned in Ref.~\cite{Aarts:2005hg}.
These characteristic properties will be confirmed also by the numerical
results presented in the next section (\ref{sec:free_lattice}).

It is important to consider physical interpretations of this
contribution.  
The term $\omega\delta(\omega)$ remains in the continuum form and 
infinite volume results \cite{Karsch:2003wy,Aarts:2005hg}, while 
it disappears at zero temperature, i.e. infinite temporal extent.
Therefore the contribution is caused by a kind of physical thermal
effect. 
The origin of the effect could be considered as a scattering with the 
thermal bath (Landau dumping) which is related to some transport 
coefficients.
From the Kubo formula, for example, a derivative of the spectral
function in the V channel, $\rho_{V}(\omega)$ which is defined by 
Eq.~(\ref{eq:spf}) using $\Gamma=\gamma_i$,
is related to the electrical conductivity $\sigma$ \cite{Gupta:2003zh}, 
\begin{eqnarray}
 \sigma = \left.\frac{1}{6}\frac{\partial}{\partial \omega}
 \rho_V(\omega)\right|_{\omega=0}.
\end{eqnarray}

Let us also note on the dilepton rate.
The dilepton rate is extracted from the spectral
functions in the vector channel of spatial 
($\rho_V(\omega)$ with $\Gamma=\gamma_i$)
and temporal ($\rho_{V_0}$ with $\Gamma=\gamma_0$) components, 
\begin{eqnarray}
\rho^{dilepton}(\omega) = \rho_V(\omega) + \rho_{V_0}(\omega).
\end{eqnarray}
$\rho_{V_0}(\omega)$ is the charge susceptibility and 
it has only the $\omega\delta(\omega)$ term
which is equivalent to that of $\rho_V(\omega)$ with the opposite sign.
Therefore the constant contribution in the dilepton rate cancels at
least for the free quark case. 

\subsection{Free quark calculations on a lattice}
\label{sec:free_lattice}
In this subsection we present numerical calculations of the meson-like 
correlators for free quarks on lattices.
The calculations are performed on 
isotropic $N_s^3\times N_t = 16^3 \times 32$ lattices, and 
the volume dependence is discussed using results for 
$96^3 \times 32$ lattice.
The free quark is described by the Wilson quark action with a bare quark
mass of $m_q=0.2$. The boundary conditions are anti-periodic in temporal
direction and periodic in spatial directions.

\begin{figure}
\resizebox{65mm}{50mm}{
 \includegraphics{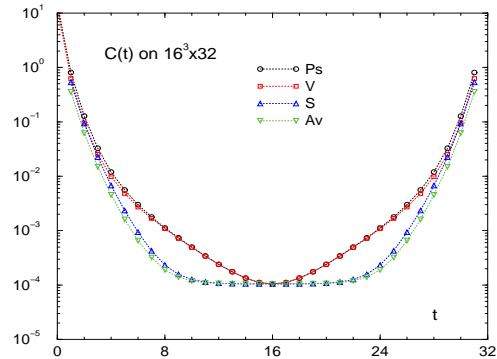}}
\caption{Meson-like correlators for each channel. 
These are free quark calculations on a $16^3\times 32$ lattice. 
The scale is logarithmic on the vertical axis.}
\label{fig:free1}
\end{figure}

Figure \ref{fig:free1} shows correlators (Eq.~(\ref{eq:corrdef})) 
for each channel.
One can clearly see a constant contribution in the S and Av channels.
The qualitative difference between the (Ps,V) and (S,Av) channels 
and the coincidence of correlators at the mid point $t=N_t/2$ can 
be explained as mentioned in the previous section.

\begin{figure}
\resizebox{65mm}{50mm}{
 \includegraphics{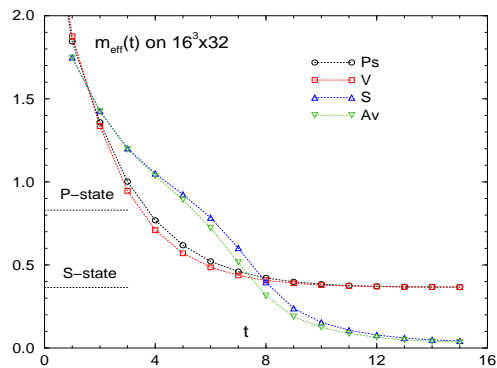}}
\caption{Effective masses of meson-like correlators for each channel.
 These are free quark calculations on a $16^3\times 32$ lattice.}
\label{fig:free2}
\end{figure}

Figure \ref{fig:free2} shows the effective masses of the correlators in
Fig.~\ref{fig:free1}.
The effective mass is defined by correlator at successive time
slices.
\begin{equation}
\frac{C(t)}{C(t+1)}=
\frac{\cosh{\left[m_{\mbox{eff}}(t)\left(\frac{N_t}{2}-t\right)\right]}}
{\cosh{\left[m_{\mbox{eff}}(t)\left(\frac{N_t}{2}-t-1\right)\right]}}
\end{equation}
The effective masses of the meson-like correlators with free quarks
should approach the energy of the two quark state without momentum for 
S-wave states (Ps and V channels) and with a minimum momentum for 
P-wave states (S and Av channels) except for the zero energy mode.
In the Wilson quark action, the free quark propagator satisfies the
dispersion relation,
\begin{eqnarray}
 \cosh{E_q(p)}=1+\frac{\sin^2(p_i)
+(m_q+\hat{p}^2/2)^2}{2(1+m_q+\hat{p}^2/2)},
\label{eq:free_disp}
\end{eqnarray}
where $\hat{p}_i=2\sin(p_i/2)$ and $E_q(p)$ is the energy of a quark
with momentum $p_i$. 
Using Eq.~(\ref{eq:free_disp}), one can calculate the lowest
energies of the two quark states in S-wave and P-wave states.
The values obtained for $N_s=16$ and $m_q=0.2$ are shown in
Fig.~\ref{fig:free2}.  
In the Ps and V channels, the zero relative momentum processes have no
constant contribution and their effective masses approach the
expected values. 
In the S and Av channels, on the other hand, their effective masses
approach the zero energy level rather than the energy of two quark
states. 

\begin{figure}
\resizebox{65mm}{50mm}{
 \includegraphics{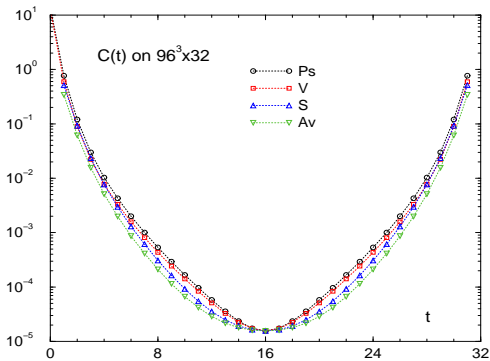}}
\caption{Meson-like correlators for each channel. 
These are free quark calculations on a $96^3\times 32$ lattice. 
The scale is logarithmic on the vertical axis.}
\label{fig:free1b}
\end{figure}

\begin{figure}
\resizebox{65mm}{50mm}{
 \includegraphics{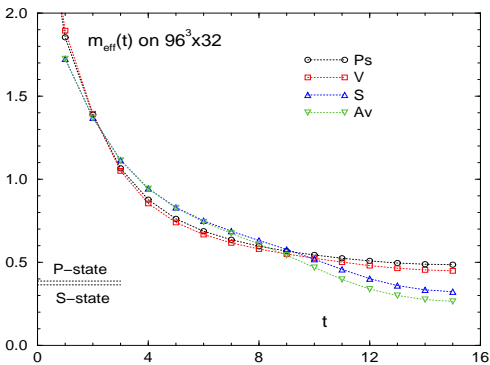}}
\caption{Effective masses of meson-like correlators for each channel.
 These are free quark calculations on a $96^3\times 32$ lattice.}
\label{fig:free2b}
\end{figure}

Next I present the results for the different volume.
Figure \ref{fig:free1b} and \ref{fig:free2b} are results of the
meson-like correlators and their effective masses on a $96^3\times 32$
lattice with the same quark mass, $m_q=0.2$, respectively.
In the latter section (\ref{sec:quench_qcd}), we present
simulations in quenched QCD at finite temperature, where
the finite temperature lattices have the similar aspect ratio,
$N_s/N_t$, and the lowest state energy in lattice units to this free
quark calculation with $N_s=96$. 
Although we find the large volume dependence between $N_s=16$ and
$N_s=96$, almost no volume dependence is seen for $N_s/N_t > 2$.
At a sufficiently large aspect ratio, we find no clear constant
contribution, but the finite constant contribution exists when the quark
mass is finite.

\subsection{An analysis to avoid the constant contribution and
  Polyakov loop sectors}
\label{sec:free_ploop}

Since the lattice QCD is formulated in Euclidean space-time,
we can isolate the lowest state contribution from correlators in the
large Euclidean time region, 
while studies of non-lowest states are not so easy.
When one want to know the information about nonzero energy states, the
constant mode makes it difficult.
The lowest scattering process yields only a constant mode in
the correlators considered in this paper, 
and does not affect the other dynamics at higher energy, 
$\omega \ge 2m_q$, at least in the free quark case.
Even in the case of interacting QCD, if the width of the peak structure
in the spectral function at $\omega = 0$ is sufficiently small,
e.g. $\omega \ll T$, the contribution yields only a constant mode in
the correlators \cite{Aarts:2002cc,Petreczky:2005nh}.
Therefore, in order to investigate the nonzero energy states in the
correlators, it is useful to remove the constant
mode from the correlators in an appropriate analysis.
In this subsection we present two analysis methods to avoid the constant
contribution from the correlators.

Of course the cosh + constant fit enables us to do so, but here we
present the methods without a fit analysis.
The constant mode in the correlators can be removed by the derivative of
the correlator with respect to $t$ \cite{Majumdar:2003xm}, where the
differential correlator $C'(t)$ is defined by a derivative of $C(t)$, 
\begin{eqnarray}
 C'(t)=C(t+1)-C(t).
\end{eqnarray}
Here $C'(t)$ is equivalent to the symmetric derivative at $t+1/2$.
One can define the effective mass of the differential correlator,
\begin{eqnarray}
\frac{C'(t)}{C'(t+1)}=
\frac{\sinh{\left[m_{\mbox{eff}}^{\mbox{diff}}(t)
\left(\frac{N_t}{2}-t\right)\right]}}
{\sinh{\left[m_{\mbox{eff}}^{\mbox{diff}}(t)
\left(\frac{N_t}{2}-t-1\right)\right]}}.
\label{eq:effmassdiff}
\end{eqnarray}
Since, however, the overlap of each state is modified by a factor of the
energy of the state in the analysis, the plateau of the effective 
masses becomes narrow.
Furthermore, in general, a derivative of an observable increases
the statistical fluctuations, thus the analysis may be difficult to
apply to actual numerical calculations. 
Therefore we adopt a rather different approach to remove the constant
contribution \footnote{We thank Frithjof Karsch for the remark on the
midpoint subtracted correlator.}. 
We define the midpoint subtracted correlator, 
\begin{eqnarray}
 \bar{C}(t)=C(t)-C(N_t/2).
\end{eqnarray}
One can also define the effective mass of the subtracted correlator,
\begin{eqnarray}
\frac{\bar{C}(t)}{\bar{C}(t+1)}=
\frac{\sinh^2{\left[\frac{1}{2}m_{\mbox{eff}}^{\mbox{sub}}(t)
\left(\frac{N_t}{2}-t\right)\right]}}
{\sinh^2{\left[\frac{1}{2}m_{\mbox{eff}}^{\mbox{sub}}(t)
\left(\frac{N_t}{2}-t-1\right)\right]}}.
\label{eq:effmasssub}
\end{eqnarray}

\begin{figure}
\resizebox{70mm}{45mm}{
 \includegraphics{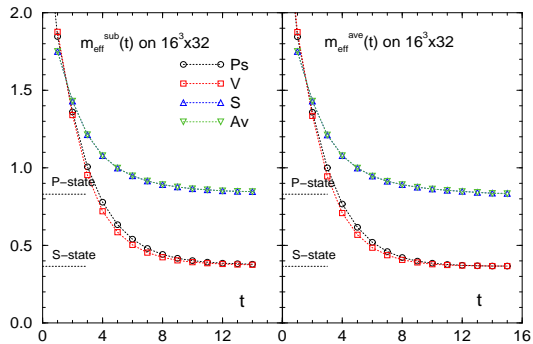}}
\caption{Effective masses for each channel
calculated from (left) the midpoint subtracted
 meson-like correlator and 
(right) the averaged correlator (averaged over the Polyakov loop
 sectors). 
These are free quark calculations on $16^3\times 32$ lattices. }
\label{fig:free4}
\end{figure}

Figure \ref{fig:free4} (left panel) shows the effective mass 
$m_{\mbox{eff}}^{\mbox{sub}}(t)$, defined in Eq.~(\ref{eq:effmasssub}), 
from the free quark results for $N_s=16$ in Fig.~\ref{fig:free1}.
The effective masses are equivalent to the usual effective mass shown 
in Fig.~\ref{fig:free2} except for the effects of the constant mode. 
The expected energies of the lowest two quark states for the S-wave and
the P-wave states can be calculated from the free quark dispersion
relation, Eq.~(\ref{eq:free_disp}). The values are 
shown in Fig.~\ref{fig:free4} as well as in Fig.~\ref{fig:free2}.
In contrast to the case of the usual effective masses in
Fig.~\ref{fig:free2}, the effective masses
$m_{\mbox{eff}}^{\mbox{sub}}(t)$ approach the expected values even
in the P-wave states.  
The analysis to avoid the constant contribution works well
at least in the free quark case.

The method using the midpoint subtracted correlators can also be applied
to studies of spectral functions $\rho_\Gamma(\omega)$ by a modification
of the kernel,   
\begin{eqnarray}
\bar{C}(t)&=&\int_0^\infty d\omega \rho_\Gamma(\omega)K^{\mbox{sub}}
(\omega,t),\\
&&K^{\mbox{sub}}(\omega,t)=
\frac{2\sinh^2\left(\frac{\omega}{2}(\frac{N_t}{2}-t)\right)}
{\sinh{\left(\omega\frac{N_t}{2}\right)}}.
\label{eq:subkernel}
\end{eqnarray}
By using the alternative kernel $K^{\mbox{sub}}(\omega,t)$, it is
possible to extract the spectral function without the
contribution in $\omega\ll T$ from e.g. the Maximum Entropy Method. 

Next we consider the second method to avoid the constant
contribution. The method utilizes the $Z_3$ transformation properties of
the correlators, where the $Z_3$ transformation is performed on the lattice
through the transformation of temporal link variables,
\begin{eqnarray}
U_4(\vec{x},t) \rightarrow e^{i2n\pi/3}U_4(\vec{x},t),  
\end{eqnarray}
where $n=0,1$ and 2, for any $\vec{x}$, at a time slice
(we adopt $t=N_t/2$).
First we consider the $Z_3$ transformation properties for the diagrams
in Fig.~\ref{fig:corr}(a) and (b).
The contribution of Fig.~\ref{fig:corr}(a) has a $Z_3$ center symmetry,
while that of Fig.~\ref{fig:corr}(b) is no longer invariant under the
$Z_3$ transformation. 
In case of the static limit, one can easily understand it by the analogy
of Polyakov loop and Wilson loop.
Even in the non-static case, the properties can be derived using a
hopping parameter expansion of the quark propagator. In any order of the
expansion, one can show this property like in the static limit.

Furthermore, for the $Z_3$ variant contribution, the $Z_3$ factor
can be factored out.
Therefore one can cancel the $Z_3$ variant contribution by averaging 
the correlators over the three sectors which are characterized by the
location of the Polyakov loop in the complex plane.
Here we define the averaged correlator $C^{\mbox{ave}}(t)$ from
$C^{p0}(t)$, $C^{p1}(t)$ and $C^{p2}(t)$, which are correlators
on the configurations with an argument of the Polyakov loop in
the complex plane fulfilling
$-\frac{2\pi}{3}\le arg( P ) \le \frac{2\pi}{3}$,
$\frac{2\pi}{3}\le arg( P ) \le \pi$ and 
$ -\pi \le arg( P ) \le -\frac{2\pi}{3}$ respectively, 
where $P$ is a value of the Polyakov loop on the given configuration,
\begin{eqnarray}
 C^{\mbox{ave}}(t)=\frac{1}{3}\left(C^{p0}(t)+C^{p1}(t)+C^{p2}(t)\right).
\label{eq:defavecorr}
\end{eqnarray}
Each correlator can be calculated on the $Z_3$ transformed gauge
configurations. 

The second method to avoid the constant contribution is in a way similar
to removing the backward quark propagations by using a cancellation
between results with periodic and anti-periodic boundary conditions in
the $t$ direction. 
However the second method is beyond a technical step to separate the
constant mode from the others.
The effect of $Z_3$ symmetrization is to remove from the partition
function all states with non-integer baryon number (whether in
quenched or full QCD).  This has been discussed, e.g., in  
Ref.~\cite{Kratochvila:2006jx}.

Here we should mention that the cancellation of the constant mode is not
exact in the second method. 
Because, e.g., in the case of a diagram with a 3 times wrapping
quark line, the contribution also has the $Z_3$ center symmetry. 
Therefore it will not be canceled in the analysis.
(This situation is also similar to the cancellation of the backward
quark propagation as mentioned above. This cancellation is also not
exact.
The remaining $Z_3$ symmetric contribution can be removed by
considering the zero-baryon partition function, as discussed as well
in Ref.~\cite{Kratochvila:2006jx}.)
However, the $Z_3$ invariant constant contributions is of 
$O(e^{-3m_qL_t})$ or less, while that for the leading (canceled)
contribution is $O(e^{-m_qL_t})$. 
Therefore the averaged correlator is useful to see
a qualitative effect of the constant contribution in the correlators,
when $m_q L_t$ is not so small.

From the averaged correlator in Eq.~(\ref{eq:defavecorr}), 
one can define the effective mass $m_{\mbox{eff}}^{\mbox{ave}}(t)$,
\begin{equation}
\frac{C^{\mbox{ave}}(t)}{C^{\mbox{ave}}(t+1)}=
\frac{\cosh{\left[m^{\mbox{ave}}_{\mbox{eff}}(t)
\left(\frac{N_t}{2}-t\right)\right]}}
{\cosh{\left[m^{\mbox{ave}}_{\mbox{eff}}(t)
\left(\frac{N_t}{2}-t-1\right)\right]}}.
\end{equation}
The effective masses $m_{\mbox{eff}}^{\mbox{ave}}(t)$ for the free quark 
case are presented in the right panel of Fig.~\ref{fig:free4}.
Similar to the effective masses from the midpoint subtracted
correlators, $m_{\mbox{eff}}^{\mbox{ave}}(t)$ approach the expected
nonzero lowest energies even for the P-wave states. 
However the method does not exactly remove the constant contribution, a
part of the constant contribution remains. 

The method using the midpoint subtracted correlators does not work for
correlators with finite momentum and non-degenerate masses of valence
quarks, e.g. heavy-light meson. Although these cases are out of the
scope of this study, the averaged correlator method works even for such
meson-like operators. 
Furthermore provided the peak structure of spectral function at
$\omega=0$ has a wide width $\omega > T$ due to interactions,
the subtracted correlator method may not even work in the case of
correlators with zero momentum and degenerate quarks, while the averaged 
correlator method can be applied even in this case. 

\section{The constant contribution in quenched lattice QCD at finite
 temperature} 
\label{sec:quench_qcd}

\subsection{Lattice setup}
\label{sec:setup}

In this section we demonstrate effect of the constant contribution and
several analyses to avoid it in quenched QCD at finite temperature.
In the simulation we calculate the charmonium correlators, which are
defined like in Sect.~\ref{sec:freecorr} at the charm quark mass.
The temporal correlators (or the spectral functions) of charmonia
are important to discuss the $J/\psi$ suppression which is considered as
one of the most important signals of QGP formation in heavy ion
collision experiments.
Several groups have investigated the charmonium correlators
in finite temperature lattice QCD
\cite{Umeda:2002vr,Datta:2003ww,Asakawa:2003re,Aarts:2006nr,Jakovac:2006sf}.  
In such studies one has to take into account the constant contribution
to the charmonium correlators (or spectral functions).
Because this contribution may not be negligible in finite temperature
lattice QCD. In fact it is a genuine effect in finite temperature QCD
as mentioned in Sect.~\ref{sec:freecorr}.

In finite temperature lattice QCD the temporal correlators are
restricted within the inverse of the temperature. In order to keep
sufficient number of data points in temporal direction  
at high temperature, we adopt an anisotropic lattice, on which the
temporal lattice spacing $a_t$ is smaller than the spatial one $a_s$.
The temperature is varied by the temporal lattice sizes $N_t$ with a
fixed temporal lattice spacing $a_t$. 
Such calculations with a fixed lattice spacing also have the
advantage to investigate thermal effects in correlators, i.e. one can
directly compare the correlators at different temperatures 
without need for information on the beta-function.
In this section physical quantities are expressed in $a_t=1$ units
when there is no comment on the units.

The gauge configurations have been generated by an standard plaquette
gauge action with a lattice gauge coupling constant, $\beta=6.10$ and a
bare anisotropy parameter $\gamma_G=3.2108$. 
The definition of the action and parameters
are the same as that adopted in Ref.~\cite{Matsufuru:2001cp}.
Lattice sizes are $20^3\times N_t$ where $N_t=160$ at $T=0$ and $N_t=32$,
26 and 20 at $T>0$. The lattice spacings are $1/a_s=2.030(13)$ GeV and
$1/a_t=8.12(5)$ GeV which are determined from the hadronic radius
$r_0=0.5$ fm.  
The physical volume size is about $(2\mbox{fm})^3$ and temperatures at
$N_t=32$, 26 and 20 are $T=0.88T_c$, $1.08T_c$ and $1.40T_c$
respectively. The temperatures have been determined by the peak position
of the Polyakov loop susceptibility from which we found the critical
temperature around $N_t=28$.
The finite temperature lattices have been generated up to 300
configurations after 20,000 sweeps for equilibration, each configuration
is separated by 500 sweeps to reduce the autocorrelation.
At zero temperature, we have 60 configurations with the same condition.
In the calculations, the error analysis is performed by the jackknife
estimation.

\begin{figure}
\resizebox{70mm}{45mm}{
 \includegraphics{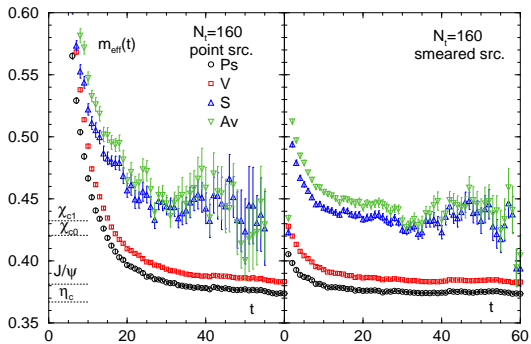}}
\caption{Effective masses from charmonium correlators in quenched QCD at
 zero temperature. The left panel is the result with the local charmonium
 operators and the right one is the result with the spatially extended
 charmonium operators. Dotted lines in the left panel show the
 experimental values for each channel.}
\label{fig:quench1}
\end{figure}

For the quark fields, we adopt an $O(a)$ improved Wilson quark action
with tadpole improved tree level clover coefficients.
Although the definition of the quark action is the
same as in Ref.~\cite{Matsufuru:2001cp}, 
we adopt a different choice of the Wilson
parameter $r=1$ to suppress effects of lattice artifacts
in higher states of charmonium \cite{Karsch:2003wy}.
Then the parameters in the quark action \footnote{Our lattice setup 
is very similar to one of Ref.~\cite{Jakovac:2006sf}. Although
our notation of the parameters are different from theirs, the
corresponding parameters are consistent with the ones they adopted.} 
are $\kappa=0.10109$, $c_E=1.911$, $c_B=3.164$ and $\gamma_F=4.94$. 
Figure~\ref{fig:quench1} shows the usual effective masses calculated
with the parameters listed above.  
The results at zero temperature satisfy a relativistic meson
dispersion relation with the same anisotropy as the 
renormalized gauge anisotropy $a_s/a_t=4$. 
The left and right panels in Fig.~\ref{fig:quench1} show the results
from the local and spatial extended operators. 
The latter operators are defined by
$O_\Gamma(\vec{x},t)=\sum_{\vec{y}}\phi(\vec{y}) 
\bar{q}(\vec{x}-\vec{y},t)\Gamma q(\vec{x},t)$ 
with a smearing function $\phi(\vec{x})$ in Coulomb gauge.
In this calculation the spatially extended operators are adopted only in 
the source operators, the sink operators are local in any cases.
The smearing function is the same as that in Ref.~\cite{Umeda:2002vr},
i.e. $\phi(\vec{x})=\exp(-A|\vec{x}|^P)$ where $A$ and $P$ are parameters
determined by a matching with the charmonium wave function as
$A=0.2275$ and $P=1.258$.

From the zero temperature calculation with the spatially extended
operators we obtain the charmonium masses in physical units,
$m_{\eta_c}=3033(19)$ MeV and $m_{J/\psi}=3107(19)$ MeV,
which are slightly heavier than their experimental values 
$m_{\eta_c}^{\mbox{exp.}}=2980$ MeV and 
$m_{J/\psi}^{\mbox{exp.}}=3097$ MeV \cite{Yao:2006px}.
The experimental values in lattice units are also shown in 
Fig.~\ref{fig:quench1}.
The results for the P-wave states are rather noisy at our current
statistics, and the actual values are not so important for this study.
Therefore we do not show the numbers in MeV, however their effective
masses show a plateau at reasonable values. 

\subsection{Finite temperature results}

\begin{figure}
\resizebox{70mm}{45mm}{
 \includegraphics{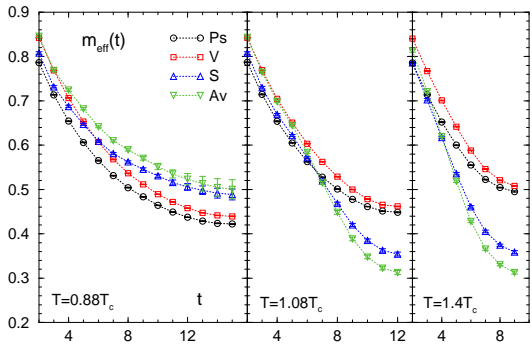}}
\caption{Effective masses from charmonium correlators
in quenched QCD at finite temperature. 
These are results with local operators for each channel.} 
\label{fig:quench2}
\end{figure}

In this section we discuss the result of the charmonium correlators at
finite temperature. The definitions for several types of effective
masses can be taken over from that of the free quark calculations in 
Sect.~\ref{sec:FreeTheory}.  
First we show the usual effective masses, $m_{\mbox{eff}}(t)$ at finite
temperature in Fig.~\ref{fig:quench2}. 
Since these are results obtained with local operators, there is no
plateau region at any temperature or channel we chose.
It can be expected from zero temperature calculation and the effective
masses show reasonable behavior below $T_c$, at which the charmonium
correlators does not show large difference from the zero temperature
results \cite{Umeda:2002vr,Datta:2003ww,Jakovac:2006sf}. 
We find rather large statistical fluctuations in the P-wave states, 
a reason for that is given later.  
Just above $T_c$, we see drastic changes in the P-wave state channels. 
The changes may cause the dissolution of $\chi_c$ states just above
$T_c$ as already reported in
Ref.~\cite{Datta:2003ww,Aarts:2006nr,Jakovac:2006sf}.
\begin{figure}
\resizebox{70mm}{45mm}{
 \includegraphics{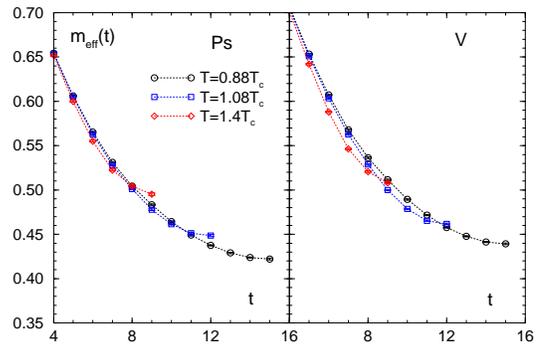}}
\caption{Temperature dependence of the effective masses from charmonium
 correlators in quenched QCD at finite temperature. 
 The left panel shows the Pseudoscalar
 (Ps) channel and the right one is the Vector (V) channel. These lowest
 states correspond to $\eta_c$ and $J/\psi$. }
\label{fig:quench3}
\end{figure}

On the other hand, the S-wave states, i.e. the Ps and V channels, show
small changes up to  $1.4T_c$.  
This temperature dependences are shown in Fig.~\ref{fig:quench3} for
each channel. When we compare the effective masses at the same time
slices, we find no large change up to $1.4T_c$ in both channels as well. 
The results are consistent with previous lattice studies of charmonium
spectral functions 
\cite{Umeda:2002vr,Datta:2003ww,Asakawa:2003re,Aarts:2006nr,Jakovac:2006sf}, 
in which the spectral functions for the Ps and V channels in the 
deconfinement phase (but at not so high temperature) show a peak
structure around the $\eta_c$ and $J/\psi$ masses like the result at
zero temperature. 

\subsection{Midpoint subtracted correlator analysis}

As discussed for free quarks in Sect.~\ref{sec:FreeTheory}, 
the constant contribution may dominate the meson correlators for some
channels in the deconfinement phase. 
In order to see the contribution in the meson correlators we calculate
the effective mass from the midpoint subtracted correlator, 
$m_{\mbox{eff}}^{\mbox{sub}}(t)$ defined in Eq.~(\ref{eq:effmasssub}).
\begin{figure}
\resizebox{70mm}{45mm}{
 \includegraphics{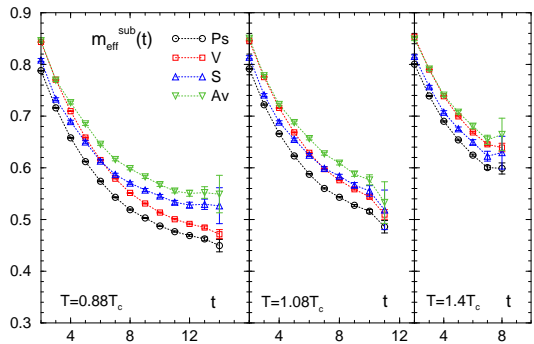}}
\caption{Effective masses from the midpoint subtracted charmonium
correlators in quenched QCD at finite temperature. 
These are results with local operators for each channel.}
\label{fig:quench4}
\end{figure}
Figure \ref{fig:quench4} shows the results of
$m_{\mbox{eff}}^{\mbox{sub}}(t)$ in quenched QCD at finite temperature. 
As we expected, the effective masses at each temperature in the S-wave
states show small differences from that of zero temperature, when
comparing them at the same time slices. 
In the P-wave states, however, the drastic changes of the usual
effective masses $m_{\mbox{eff}}(t)$ are absent in the effective masses
from the midpoint subtracted correlators $m_{\mbox{eff}}^{\mbox{sub}}(t)$.
Furthermore the similar behaviors of the effective masses hold till
$1.4T_c$ as in the case of the S-wave states.
The results indicate that the drastic changes of the usual correlators
(and effective masses) are caused only by the constant contribution to
the correlators. 
The situation is very similar to the free quark case discussed in
Sect.~\ref{sec:FreeTheory}.

\subsection{Polyakov loop sector dependence of correlators}
\label{sec:quench_ploop}

\begin{figure}
\resizebox{70mm}{45mm}{
 \includegraphics{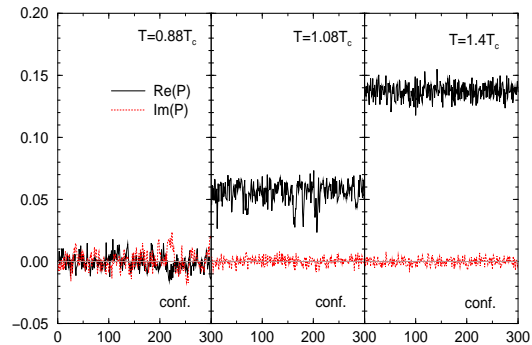}}
\caption{Time histories of the real and imaginary part of the Polyakov
 loop $P$ at each temperature simulation.}
\label{fig:quench5}
\end{figure}

Next we discuss the Polyakov loop sector dependence of meson correlators 
as mentioned in Sect.~\ref{sec:free_ploop}.
First of all we show time histories of the Polyakov loop on our finite 
temperature configurations in Fig.~\ref{fig:quench5}. 
For all configurations above $T_c$ the Polyakov loop lies in the real
sector. 

\begin{figure}
\resizebox{70mm}{45mm}{
\includegraphics{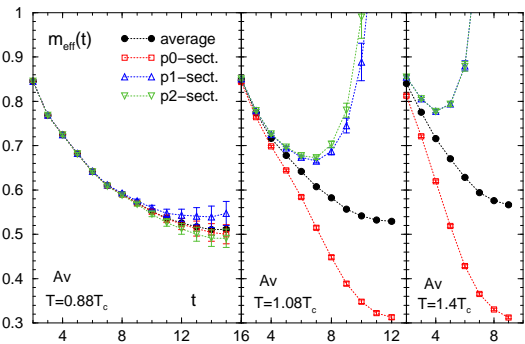}}
\caption{Polyakov loop sector dependence of Effective masses for the Av
channel. p0, p1 and p2 sector means the result has been obtained on
configurations with the Polyakov loop restricted to the real and
(positive and negative) imaginary sectors respectively. 
Average means averaged over all sectors as in Eq.~(\ref{eq:defavecorr}).} 
\label{fig:quench6}
\end{figure}

Thus Fig.~\ref{fig:quench2} gives the result on configurations with
the Polyakov loop being in the real sector. 
The correlators for the imaginary sectors are calculated
on configurations generated by $Z_3$ transformations from that in the
real sector. As an example for the imaginary sectors we show the
effective masses calculated from $C^{p0}(t)$, $C^{p1}(t)$, $C^{p2}(t)$
and $C^{\mbox{ave}}(t)$, which are defined in
Sect.~\ref{sec:free_ploop}, in Fig.\ref{fig:quench6}. 
As discussed in Sect.~\ref{sec:free_ploop}, the $Z_3$ transformation
changes a part of the constant contribution by a factor of $Z_3$, i.e.
$e^{i2n\pi/3}$ where $n=0,1$ and 2.
Since our definition of the meson correlators is real, the factor acts 
like $\mbox{Re}\left(e^{i2n\pi/3}\right)$.
It means that the constant contributions in the imaginary sectors become 
negative when the value in the real sector is positive.
Hence the correlators of the P-wave states approach a positive constant
value like in the real sector, as shown in Fig.\ref{fig:free1}, while
the correlators for the imaginary sectors approach a negative value.  
As a result the effective masses of the P-wave states diverge 
at a time slice as shown in Fig.~\ref{fig:quench6}. 

Here we comment on the large statistical fluctuations of correlators for 
the P-wave states just below $T_c$ as can be seen in
Fig.~\ref{fig:quench2}. 
Below $T_c$, in principle, the constant contribution vanishes due to
the infinite energy of a single quark state in the confinement phase.
However the contribution on each configuration may have a small value
near $T_c$ even in confinement phase, like the Polyakov loop just below
$T_c$ in Fig.~\ref{fig:quench5}. 
In this case the effect of the constant contribution appears in
the correlators. 
Since the Polyakov loop sector frequently changes below $T_c$, 
the effects appear as the large fluctuation in the correlators. 

When the averaged correlator defined in Eq.~(\ref{eq:defavecorr}) is
calculated, one can expect that most of the constant contributions are 
canceled like in the free quark case (Sect.~\ref{sec:free_ploop}).
\begin{figure}
\resizebox{70mm}{45mm}{
 \includegraphics{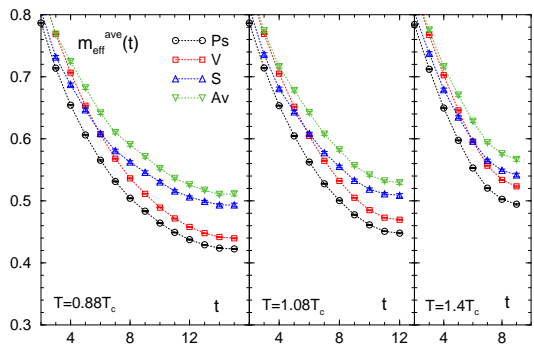}}
\caption{Effective masses from the averaged charmonium
correlators (averaged over the three sectors of the Polyakov loop) in
quenched QCD at finite temperature. 
Configurations with the Polyakov loop in the imaginary sectors have been
obtained by a $Z_3$ transformation of the corresponding real sector
configuration.
These are results with local operators for each channel.}
\label{fig:quench7}
\end{figure}
Figure \ref{fig:quench7} shows the results of 
$m_{\mbox{eff}}^{\mbox{ave}}(t)$ at each temperature.
We find almost the same behavior as in Fig.~\ref{fig:quench4}.
The results also suggest that the drastic changes in the P-wave states
just above $T_c$ are coming from the constant contribution only. 

\subsection{Results with spatially extended operators}

Finally we present results of meson correlators with spatially extended
operators.
The spatially extended operators are constructed with a smearing function
which yields a spatial distribution of quark (and anti-quark) source(s).
The distribution also changes the momentum distribution of the quark
propagation.
For example, a point source function yields any (lattice) momenta with 
equal distribution, while a wall source function yields only quark
propagation with zero momentum.
In the case of our source the (smearing) function provides quark
propagation with any momenta but quark propagators with small momentum
are enhanced. 
Therefore the correlators with spatially extended operators defined, as
in Sect.~\ref{sec:setup}, yield (relatively) large overlap with the
lower energy states.

\begin{figure}
\resizebox{70mm}{45mm}{
 \includegraphics{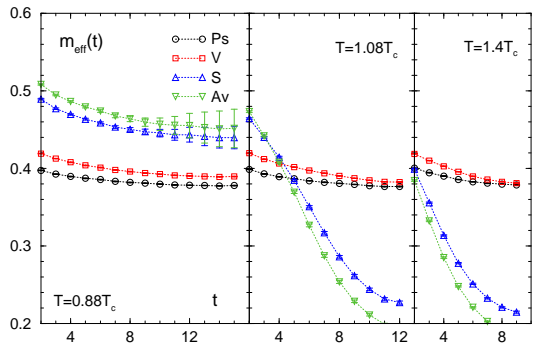}}
\caption{The same as in Fig.~\ref{fig:quench2} but 
for spatially extended operators in each channel.} 
\label{fig:quench8}
\end{figure}
Figure \ref{fig:quench8} is the result for the usual effective masses
with the spatially extended operators. Other conditions are the same as 
in Fig.~\ref{fig:quench2}.
Below $T_c$, in contrast to the case of local operators the effective
masses reach a plateau due to the larger overlap with the lowest state.
Above $T_c$ the effective masses of the P-wave states
change more than in the case of local operators. 
Because the constant contribution is enhanced by the smearing function 
more than the other contributions in the case.
The effect of the constant contribution is also present in the V
channel.  
At least in the free quark case, as we discussed in
Sect.~\ref{sec:freecorr}, the constant contribution to the V channel is 
smaller than that to the P-wave states for finite quark masses and small
volumes. 
In fact, even in quenched QCD, it is difficult to see the
effect in the V channel in local operators. 
However we find small but finite changes even in the V channel above
$T_c$ while no changes are seen in the Ps channel as we expected. 

\begin{figure}
\resizebox{70mm}{45mm}{
 \includegraphics{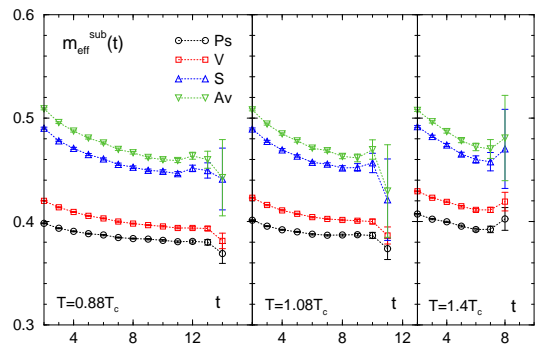}}
\caption{The same as in Fig.~\ref{fig:quench4} but 
for spatially extended operators in each channel.} 
\label{fig:quench9}
\end{figure}
We can say that the small changes in the V channel is caused by the
constant contribution, because the analysis with the midpoint subtracted 
correlators shows almost no changes in the V channel above $T_c$ as it
can be seen in Fig.~\ref{fig:quench9}. Similar to the results with local
operators, the effective masses of the P-wave states also show small
change at any temperatures.

\begin{figure}
\resizebox{70mm}{45mm}{
 \includegraphics{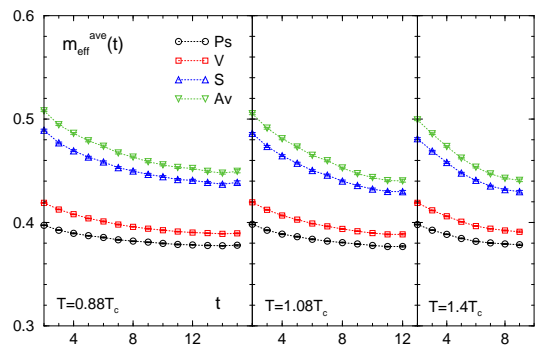}}
\caption{The same as in Fig.~\ref{fig:quench7} but 
for spatially extended operators in each channel.} 
\label{fig:quench10}
\end{figure}
Figure \ref{fig:quench10} is the same figure as Fig.~\ref{fig:quench7}, 
but with spatially extended operators.
We find similar behavior for both cases of operators.

\section{Discussion}
\label{sec:Discussion}

\begin{figure}
\resizebox{70mm}{45mm}{
 \includegraphics{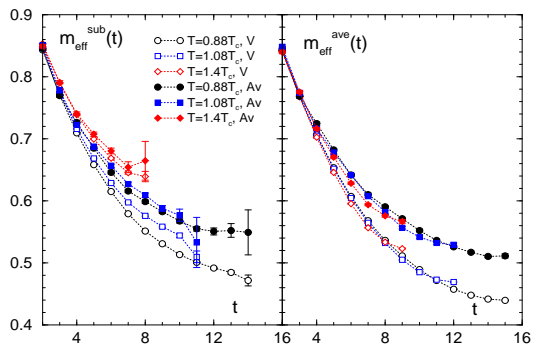}}
\caption{Temperature dependences of the effective masses from the
midpoint subtracted charmonium correlators (left) 
and the averaged charmonium correlators (averaged over the three sectors
of the Polyakov loop) (right) in quenched QCD at finite temperature. 
These are results with local operators of the V and the Av channels, 
whose lowest states correspond to the $J/\psi$ and $\chi_{c1}$ states.}
\label{fig:quench11}
\end{figure}

\begin{figure}
\resizebox{70mm}{45mm}{
 \includegraphics{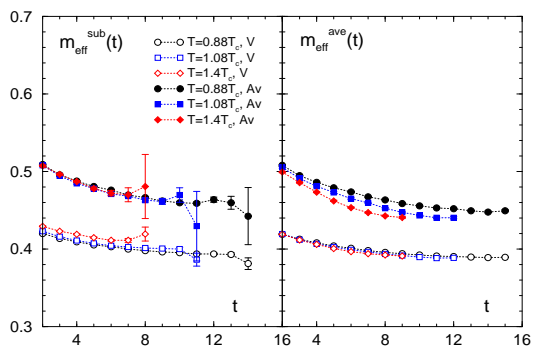}}
\caption{The same as in Fig.~\ref{fig:quench11} but 
for spatially extended operators in the V and the Av channels.} 
\label{fig:quench12}
\end{figure}

Let us discuss details on $\chi_c$ states using the results of the
previous sections. In this study we find that the drastic changes of the
(usual) correlators in the P-wave states are caused by the constant
contribution,
while the correlators without the constant mode yield small change
till, at least, $1.4T_c$ even in the P-wave states. 
The changes are of similar size as that of the
S-wave states as one can see in Fig.~\ref{fig:quench11} and 
Fig.~\ref{fig:quench12}.

In the case of the spatially extended operators, the averaged effective
mass $m_{\mbox{eff}}^{\mbox{ave}}(t)$ for the Av channel gradually
reduces as the temperature increases. 
The change seems to be effected by the constant contributions with 
the $Z_3$ center symmetry as discussed in Sect.~\ref{sec:free_ploop}.
In fact, we can not find any hints of such a change in
$m_{\mbox{eff}}^{\mbox{sub}}(t)$ in the left panel of
Fig.~\ref{fig:quench12}. Furthermore it is reasonable to find the effect
only in case of the spatially extended operators, because the $Z_3$
symmetric constant contribution is enhanced by the smearing
function. 

Although the constant mode coming from the scattering process is
important to investigate some transport phenomena, it does not affect 
the spectral function in the $\omega \gg T$ region for the charmonium 
correlators with zero spatial momentum.
Therefore the dissolution of $\chi_c$ states just above $T_c$ as
discussed in e.g. Ref.~\cite{Datta:2003ww,Aarts:2006nr,Jakovac:2006sf}
might be misleading. 
When there is the constant contribution, the hadronic state is no 
longer the lowest state. 
The situation provides some difficulties in the analysis
to reconstruct the spectral function in the higher energy part 
$\omega \gg T$ even if the MEM is applied \cite{Umeda:2005nw}. 
The difficulties may be understood by the fact that most studies of
the charmonium spectral function can not reproduce the excited states
correctly especially at finite temperature (even below $T_c$).
Furthermore, in general, correlators for the P-wave states are noisier 
than that for the S-wave states as one can see in
Fig.~\ref{fig:quench1}.  

In principle the MEM analysis can extract the correct spectral function
even if the constant contribution exists, however it is difficult to
reproduce the non-lowest part of the spectral function correctly by the
MEM analysis at finite temperature with typical statistics.
In order to investigate the properties of $\chi_c$ states above $T_c$, a
careful analysis taking the constant contribution into account is
necessary. 
A correct MEM analysis of the correlator with and without the
constant contribution should give the same results, up to the 
low energy region $\omega\ll T$.
Such a comparison can be easily performed by the MEM analysis with
the alternative kernel of Eq.~(\ref{eq:subkernel}) for midpoint
subtracted correlators. 
Failure to do so signals unreliable MEM results.

We also find a Polyakov loop sector dependence of the constant
contribution in quenched QCD. 
To study transport coefficients from the behavior of
spectral function at $\omega=0$ one has to pay attention to the
Polyakov loop sector especially in quenched QCD calculations.
From a practical viewpoint, the dependence on the Polyakov loop sector
is useful to estimate the constant contribution as discussed in
Sect.~\ref{sec:free_ploop} and Sect.~\ref{sec:quench_ploop}.

The constant contribution may also explain some other results of
previous studies of temporal correlators. 
First, in the study of the spectral function at finite temperature 
by the MEM analysis, one sometimes find a kind of divergence of the
spectral function near $\omega=0$, it may be caused by the constant
contribution. Of course it may be a simple failure of the analysis,
however the contribution near $\omega=0$ ought to exist except for 
the Ps channel. 
Secondly, in the S-wave states of charmonium, several groups have
reported that the correlator for the Ps channel shows almost no change
above the deconfinement transition, while that of the V channel shows
a small but significant change 
\cite{Umeda:2002vr,Jakovac:2006sf,Iida:2006mv}.
This can also be seen from our calculation in 
Sect.~\ref{sec:quench_qcd}.
It can be explained by the constant contribution as we did in this
paper. 
Thirdly, when one considers a spatial boundary condition dependence of
the charmonium spectrum at finite temperature as in 
Ref.~\cite{Iida:2006mv}, the P-wave states (in the free quark case) have
the lowest energy using a ``hybrid boundary'', which is an anti-periodic 
boundary in one spatial direction and periodic in the other directions.
On the other hand, the P-wave state has a larger energy with periodic
boundaries in any spatial directions. 
However the (usual) effective mass of the P-wave state correlators for
free quark calculations provides a smaller value in periodic boundary
conditions than in the hybrid boundary case. 
This unexpected behavior of the effective mass is also explained by the
scattering processes because of suppression of the scattering
contribution by the anti-periodic boundary in a spatial direction.  
Finally, the Polyakov loop sector dependence of the chiral order
parameter in quenched QCD \cite{Chandrasekharan:1998yx} may be also
explained by the effect of ($Z_3$ variant) wrapped contributions in the
chiral condensate. 

As we discussed in the last section,
the constant contribution affects the temporal correlators at finite
temperature in various cases. Therefore it is important to take this
contribution into account for studies of temporal correlators in lattice 
QCD at finite temperature.

\section*{Acknowledgments}

The author thanks F.~Karsch, S.~Sasaki, S.~Datta, T.~Yamazaki,
T.~Doi, H.~Matusufuru, P.~Petreczky, S.~Aoki and C.~Schmidt 
for helpful discussions and comments.
This work has been authored under contract number
DE-AC02-98CH1-886 with the U.S. Department of Energy.
The simulations have been performed on supercomputers (NEC SX-5) at
the Research Center for Nuclear Physics (RCNP) at Osaka University and
(NEC SX-8) at the Yukawa Institute for Theoretical Physics (YITP) at
Kyoto University.


\begin{thebibliography}{99}

\bibitem{Karsch:2001uw}
F.~Karsch, E.~Laermann, P.~Petreczky, S.~Stickan and I.~Wetzorke,
%``A lattice calculation of thermal dilepton rates,''
Phys.\ Lett.\ B {\bf 530}, 147 (2002)
[arXiv:hep-lat/0110208].
%%CITATION = HEP-LAT 0110208;%%

\bibitem{deForcrand:2000jx}
P.~de Forcrand {\it et al.}  [QCD-TARO Collaboration],
%``Meson correlators in finite temperature lattice QCD,''
Phys.\ Rev.\ D {\bf 63}, 054501 (2001)
[arXiv:hep-lat/0008005].
%%CITATION = HEP-LAT 0008005;%%

\bibitem{Umeda:2000ym}
T.~Umeda, R.~Katayama, O.~Miyamura and H.~Matsufuru,
%``Study of charmonia near the deconfining transition on an anisotropic
%lattice with O(a) improved quark action,''
Int.\ J.\ Mod.\ Phys.\ A {\bf 16}, 2215 (2001)
[arXiv:hep-lat/0011085].
%%CITATION = HEP-LAT 0011085;%%

\bibitem{Umeda:2002vr}
T.~Umeda, K.~Nomura and H.~Matsufuru,
%``Charmonium at finite temperature in quenched lattice QCD,''
Eur.\ Phys.\ J.\ C {\bf 39S1}, 9 (2005)
[arXiv:hep-lat/0211003].
%%CITATION = HEP-LAT 0211003;%%

\bibitem{Datta:2003ww}
S.~Datta, F.~Karsch, P.~Petreczky and I.~Wetzorke,
%``Behavior of charmonium systems after deconfinement,''
Phys.\ Rev.\ D {\bf 69}, 094507 (2004)
[arXiv:hep-lat/0312037].
%%CITATION = HEP-LAT 0312037;%%

\bibitem{Asakawa:2003re}
M.~Asakawa and T.~Hatsuda,
%``J/psi and eta/c in the deconfined plasma from lattice QCD,''
Phys.\ Rev.\ Lett.\  {\bf 92}, 012001 (2004)
[arXiv:hep-lat/0308034].
%%CITATION = HEP-LAT 0308034;%%

\bibitem{Aarts:2006nr}
G.~Aarts, C.~R.~Allton, R.~Morrin, A.~P.~O.~Cais, M.~B.~Oktay,
M.~J.~Peardon and J.~I.~Skullerud, 
%``Charmonium spectral functions in N(f) = 2 QCD at high temperature,''
PoS {\bf LAT2006}, 126 (2006)
[arXiv:hep-lat/0610065].
%%CITATION = HEP-LAT 0610065;%%

\bibitem{Jakovac:2006sf}
A.~Jakovac, P.~Petreczky, K.~Petrov and A.~Velytsky,
%``Quarkonium correlators and spectral functions at zero and finite
%temperature,''
arXiv:hep-lat/0611017.
%%CITATION = HEP-LAT 0611017;%%

\bibitem{Ishii:2002ww}
N.~Ishii, H.~Suganuma and H.~Matsufuru,
%``Glueball properties at finite temperature in SU(3) anisotropic lattice
%QCD,''
Phys.\ Rev.\ D {\bf 66}, 094506 (2002)
[arXiv:hep-lat/0206020].
%%CITATION = HEP-LAT 0206020;%%

\bibitem{Aarts:2002cc}
G.~Aarts and J.~M.~Martinez Resco,
%``Transport coefficients, spectral functions and the lattice,''
JHEP {\bf 0204}, 053 (2002)
[arXiv:hep-ph/0203177].
%%CITATION = HEP-PH 0203177;%%


\bibitem{Gupta:2003zh}
S.~Gupta,
%``The electrical conductivity and soft photon emissivity of the QCD plasma,''
Phys.\ Lett.\ B {\bf 597}, 57 (2004)
[arXiv:hep-lat/0301006].
%%CITATION = HEP-LAT 0301006;%%

\bibitem{Nakamura:2004sy}
A.~Nakamura and S.~Sakai,
%``Transport coefficients of gluon plasma,''
Phys.\ Rev.\ Lett.\  {\bf 94}, 072305 (2005)
[arXiv:hep-lat/0406009].
%%CITATION = HEP-LAT 0406009;%%

\bibitem{Hashimoto:1986nn}
T.~Hashimoto, K.~Hirose, T.~Kanki and O.~Miyamura,
%``MASS SHIFT OF CHARMONIUM NEAR DECONFINING TEMPERATURE AND POSSIBLE
%DETECTION IN LEPTON PAIR PRODUCTION,''
Phys.\ Rev.\ Lett.\  {\bf 57}, 2123 (1986).
%%CITATION = PRLTA,57,2123;%%

\bibitem{Matsui:1986dk}
T.~Matsui and H.~Satz,
%``J / psi SUPPRESSION BY QUARK - GLUON PLASMA FORMATION,''
Phys.\ Lett.\ B {\bf 178}, 416 (1986).
%%CITATION = PHLTA,B178,416;%%

\bibitem{Karsch:2005nk}
F.~Karsch, D.~Kharzeev and H.~Satz,
%``Sequential charmonium dissociation,''
Phys.\ Lett.\ B {\bf 637}, 75 (2006)
[arXiv:hep-ph/0512239].
%%CITATION = HEP-PH 0512239;%%

\bibitem{Digal:2001bh}
S.~Digal, P.~Petreczky and H.~Satz,
%``Sequential quarkonium suppression,''
arXiv:hep-ph/0110406.
%%CITATION = HEP-PH 0110406;%%

\bibitem{Antoniazzi:1992iv}
L.~Antoniazzi {\it et al.}  [E705 Collaboration],
%``Production of J / Psi via psi-prime and xi decay in 300-GeV/c proton and
%pi+- nucleon interactions,''
Phys.\ Rev.\ Lett.\  {\bf 70}, 383 (1993).
%%CITATION = PRLTA,70,383;%%

\bibitem{Kim:2003xt}
C.~Kim,
%``I = 2 pi pi scattering using G-parity boundary condition,''
Nucl.\ Phys.\ Proc.\ Suppl.\  {\bf 129}, 197 (2004)
[arXiv:hep-lat/0311003].
%%CITATION = HEP-LAT 0311003;%%

\bibitem{Takahashi:2005uk}
T.~T.~Takahashi, T.~Umeda, T.~Onogi and T.~Kunihiro,
%``Search for the possible S = +1 pentaquark states in quenched lattice
%QCD,''
Phys.\ Rev.\ D {\bf 71}, 114509 (2005)
[arXiv:hep-lat/0503019].
%%CITATION = HEP-LAT 0503019;%%

\bibitem{Aarts:2005hg}
G.~Aarts and J.~M.~Martinez Resco,
%``Continuum and lattice meson spectral functions at nonzero momentum and
%high temperature,''
Nucl.\ Phys.\ B {\bf 726}, 93 (2005)
[arXiv:hep-lat/0507004].
%%CITATION = HEP-LAT 0507004;%%

\bibitem{Karsch:2003wy}
F.~Karsch, E.~Laermann, P.~Petreczky and S.~Stickan,
%``Infinite temperature limit of meson spectral functions calculated on  the
%lattice,''
Phys.\ Rev.\ D {\bf 68}, 014504 (2003)
[arXiv:hep-lat/0303017].
%%CITATION = HEP-LAT 0303017;%%

%\bibitem{Carpenter:1984dd}
%D.~B.~Carpenter and C.~F.~Baillie,
%%``Free Fermion Propagators And Lattice Finite Size Effects,''
%Nucl.\ Phys.\ B {\bf 260}, 103 (1985).
%%%CITATION = NUPHA,B260,103;%%

\bibitem{Petreczky:2005nh}
P.~Petreczky and D.~Teaney,
%``Heavy quark diffusion from the lattice,''
Phys.\ Rev.\ D {\bf 73}, 014508 (2006)
[arXiv:hep-ph/0507318].
%%CITATION = HEP-PH 0507318;%%

\bibitem{Majumdar:2003xm}
P.~Majumdar, Y.~Koma and M.~Koma,
Nucl.\ Phys.\  B {\bf 677}, 273 (2004)
[arXiv:hep-lat/0309003].
%%CITATION = NUPHA,B677,273;%%

\bibitem{Kratochvila:2006jx}
S.~Kratochvila and P.~de Forcrand,
Phys.\ Rev.\  D {\bf 73}, 114512 (2006)
[arXiv:hep-lat/0602005].
%%CITATION = PHRVA,D73,114512;%%

\bibitem{Matsufuru:2001cp}
H.~Matsufuru, T.~Onogi and T.~Umeda,
%``Numerical study of O(a) improved Wilson quark action on anisotropic
%lattice,''
Phys.\ Rev.\ D {\bf 64}, 114503 (2001)
[arXiv:hep-lat/0107001].
%%CITATION = HEP-LAT 0107001;%%

\bibitem{Yao:2006px}
W.~M.~Yao {\it et al.}  [Particle Data Group],
%``Review of particle physics,''
J.\ Phys.\ G {\bf 33} (2006) 1.
%%CITATION = JPHGB,G33,1;%%

\bibitem{Umeda:2005nw}
T.~Umeda and H.~Matsufuru,
%``Remarks on the maximum entropy method applied to finite temperature lattice
%QCD,''
PoS {\bf LAT2005}, 154 (2006)
[arXiv:hep-lat/0510026].
%%CITATION = HEP-LAT 0510026;%%

\bibitem{Iida:2006mv}
H.~Iida, T.~Doi, N.~Ishii, H.~Suganuma and K.~Tsumura,
%``Charmonium properties in deconfinement phase in anisotropic lattice  QCD,''
Phys.\ Rev.\ D {\bf 74}, 074502 (2006)
[arXiv:hep-lat/0602008].
%%CITATION = HEP-LAT 0602008;%%

\bibitem{Chandrasekharan:1998yx}
S.~Chandrasekharan, D.~Chen, N.~H.~Christ, W.~J.~Lee, R.~Mawhinney and
	P.~M.~Vranas, 
%``Anomalous chiral symmetry breaking above the QCD phase transition,''
Phys.\ Rev.\ Lett.\  {\bf 82}, 2463 (1999)
[arXiv:hep-lat/9807018].
%%CITATION = HEP-LAT 9807018;%%

\end{thebibliography}
\end{document}